\newcommand{\ee}{\end{equation}}
\newcommand{\bea}{\begin{eqnarray}}
\newcommand{\eea}{\end{eqnarray}}
\newcommand{\bes}{\begin{subequations}}
\newcommand{\ees}{\end{subequations}}
\newcommand{\beg}{\begin{gather}}
\newcommand{\eeg}{\end{gather}}

\newcommand{\br}{{\bf r}}

\newcommand{\kp}{k^{\prime}}
\newcommand{\ip}{i^{\prime}}

\newcommand{\bei}{\begin{itemize}}
\newcommand{\ei}{\end{itemize}}

\newcommand{\bsh}{\begin{shadebox}}
\newcommand{\esh}{\end{shadebox}}
\newcommand{\besh}{\mbox{ }\\ \begin{shadebox}}
\newcommand{\eesh}{\end{shadebox}\\ \mbox{ }}
\newcommand{\non}{\nonumber}

\newcommand{\bel}{\begin{list}{$\bullet$}{
\setlength{\itemsep}{0pt}
\setlength{\topsep}{5pt}}}
\newcommand{\el}{\end{list}}

\newcommand{\bec}{\begin{center}}
\newcommand{\eec}{\end{center}}
\newcommand{\ben}{\begin{enumerate}}
\newcommand{\een}{\end{enumerate}}

\newcommand{\eff}{e\!f\!f}

\def\dint12{\int \!\!d\br_1 \!\!\int \!\!d\br_2~}
\def\dpint12{\int \!\!d\br_1^{\prime} \!\!\int \!\!d\br_2^{\prime}~}

\makeatletter

\def\eqnarray{
  \stepcounter{equation}
  \let\@currentlabel=\theequation
  \global\@eqnswtrue
  \global\@eqcnt\z@
  \tabskip\@centering
  \let\\=\@eqncr
  $$\halign to \displaywidth\bgroup\@eqnsel\hskip\@centering
  $\displaystyle\tabskip\z@{##}$&\global\@eqcnt\@ne
  \hfill$\displaystyle{{}##{}}$\hfill
  &\global\@eqcnt\tw@$\displaystyle\tabskip\z@{##}$\hfill
  \tabskip\@centering&\llap{##}\tabskip\z@\cr}

\@addtoreset{equation}{section}
\renewcommand{\theequation}{\arabic{section}.\arabic{equation}}

\makeatother

\documentclass[byrevtex,pra,floatfix,nobibnotes,nofootinbib,10pt]{revtex4}

\usepackage{graphicx}
\usepackage{amsmath}

\makeatletter
\renewcommand{\theequation}{%
\thesection.\arabic{equation}} 
\@addtoreset{equation}{section} 
\makeatother

\begin{document}

\vspace*{3cm}

\title{
Rheological Study of Transient Networks with \\
Junctions of Limited Multiplicity II. \\
Sol/Gel Transition and Rheology
\\
}

\author{Tsutomu~Indei}
\email{indei@fukui.kyoto-u.ac.jp}
\affiliation{
Fukui Institute for Fundamental Chemistry,
Kyoto University, Kyoto 606-8103, Japan \\
{\rm Tel:~+81-75-711-7894 \\
Fax:~+81-75-781-4757\\
}}

\maketitle

%%%%%%%%%%%%%%%%%%%%%%%%%%%%%%%%%%%%%%%%%%%%%%%%%%%%%%%%%%%%%%%%%%%%%%%%%%%%%
%%%%%%%%%%%%%%%%%%%%%%%%%%%%%%%%%%%%%%%%%%%%%%%%%%%%%%%%%%%%%%%%%%%%%%%%%%%%%
%%%%%%%%%%%%%%%%%%%%%%%%%%%%%%%%%%%%%%%%%%%%%%%%%%%%%%%%%%%%%%%%%%%%%%%%%%%%%

\bec
{\large Abstract}
\eec

Viscoelastic and thermodynamic properties of transient gels
formed by telechelic polymers are studied
on the basis of the transient network theory
that takes account of the correlation among polymer chains via network junctions.
The global information of the gel is incorporated into the theory
by introducing the elastically effective chains 
according to the criterion by Scanlan and Case.
We also consider effects of superbridges whose backbone is formed by 
several chains connected in series with several breakable junctions inside.
Near the critical concentration for the sol/gel transition,
superbridges becomes infinitely long along the backbone,
thereby leading to the short relaxation time $\tau$ of the network.
It is shown that $\tau$ is proportional to the 
concentration deviation $\Delta$ near the gelation point.
The plateau modulus $G_{\infty}$ increases as the cube of $\Delta$
near the gelation point as a result of the mean-field treatment,
and hence the zero-shear viscosity increases as $\eta_0\sim G_{\infty}\tau\sim\Delta^4$.
The dynamic shear moduli are well described in terms of the Maxwell model,
and it is shown that 
the present model can explain the 
concentration dependence 
of the dynamic moduli
for aqueous solutions of telechelic poly(ethylene oxide).

%
%their concentration dependence
%agree 
%%well 
%with experimental data from literature 
%for aqueous solutions of telechelic poly(ethylene oxide).

\vspace{.5cm}

%%%%%%%%%%%%%%%%%%%%

\vspace*{3cm}

\bec

\vspace*{1cm}

\noindent
%thermoreversible gel;~
associating polymers;~
transient network theory;~
junction multiplicity;~
sol/gel transition;~
Scanlan-Case criterion;~
superbridges;~

\newpage

%%%%%%%%%%%%%%%%%%%%%%%%%%%%%%%%%%%%%%%%%%%%%%%%%%%%%%%%%%%%%%%%%%%%%%%%%%%%%
%%%%%%%%%%%%%%%%%%%%%%%%%%%%%%%%%%%%%%%%%%%%%%%%%%%%%%%%%%%%%%%%%%%%%%%%%%%%%
%%%%%%%%%%%%%%%%%%%%%%%%%%%%%%%%%%%%%%%%%%%%%%%%%%%%%%%%%%%%%%%%%%%%%%%%%%%%

\section{Introduction}

Transient gels formed by associating polymers 
have been attracted widespread interests in recent years. \cite{winyek}
Associating polymers are polymer chains carrying specific groups 
capable of forming aggregates through noncovalent 
bonding.\cite{annable1,jen2,winnik,fran1,pham0,pham,serero0,serero,kujawa,meng}
Above a certain polymer concentration, they form a transient gel
by connecting sticky groups on polymers.
This transformation is thermoreversible in general.
In the previous paper\cite{in1} (referred to as I in the following), 
we presented a theoretical model of transient gels formed by 
junctions comprised of limited number of hydrophobic groups 
with an intention to understand thermodynamic properties of
linear rheology of telechelic associating polymer systems.
As the first attempt, elastically effective chains (or active chains)
were defined locally, i.e., chains whose both ends are connected to other 
end groups are elastically effective
irrespective of whether these groups are incorporated into
an infinite network (gel) or not.
We could explain, to some extent, the concentration dependence of 
the dynamic shear moduli described in terms of the Maxwell model,
but the sol/gel transition of the system could not be treated properly
due to the local definition of active chains.

In this paper, we take account of
the global information of the infinite network by making use of 
the criterion suggested by Scanlan \cite{scanlan} and Case \cite{case}
for a chain to be active. 
This criterion states that telechelic chains are elastically effective
if their both ends are connected to
junctions with at least three paths to the infinite network.
We assume that these chains deform according
to the macroscopic deformations applied to the gel.
Static properties of transient gels have been studied by Tanaka and Ishida 
on the basis of this criterion.\cite{tanaishi}
We here consider not only primary active chains 
(referred to as primary bridges in this paper)
but also active superchains (called superbridges)
whose backbone is an aggregate of several bridges connected in series.
Effects of superbridges cannot be negligible 
especially when one study dynamical properties of transient gels because they
enhance the relaxation time of the network
due to a number of internal junctions possible to break
as suggested by Annable {\it et al}..\cite{annable1}
We can describe the transition between the sol state and 
gel state (appearance of the infinite network) in this theoretical framework.
The critical behavior of viscoelastic quantities 
near the sol/gel transition point is shown to be much affected by superbridges.

It has been established up to now
that telechelic polymers self-assemble 
in dilute solution to form flowerlike micelles.
%%%%%%%%%%%%%%%%%%%%%%%%%%%%%%%%%%%%%%%%%%%%%%%%%%
%
%  Semenov理論を紹介すると，混乱するのでやめよう．
% 
%Semenov {\it et al.} \cite{semenov} 
%extended the Daoud-Cotton model for star polymers 
%%by taking account of
%so that it includes the possibility of bridging
%that causes the attraction between micelles.
%%%%%%%%%%%%%%%%%%%%%%%%%%%%%%%%%%%%%%%%%%%%%%%%%%%
%%
%Semenov {\it et al.} \cite{semenov} 
%extended the Daoud-Cotton model for star polymers 
%and treated flower micelles
%by taking account of the possibility of bridging
%that causes the attraction between micelles.
%
Pham {\it et al.} \cite{pham0,pham} indicated that 
the solution of flowerlike micelles resembles
a colloidal dispersion of adhesive hard spheres
in the concentration dependence of the shear modulus.
Quite recently, Meng and Russel \cite{meng} showed that
the colloidal theory describing the nonequilibrium structure of dispersions 
under shear explains the high-frequency plateau modulus
of telechelic poly(ethylene oxide) (PEO).
In this paper, we attempt to theoretically describe 
linear rheology of telechelic polymers
in the absence of intramolecular associations.
It can be shown that experimentally observed dynamic shear moduli
that are characterized by the high-frequency plateau modulus, 
zero-shear viscosity and relaxation time,
are well described in terms of this theoretical treatment.
%%%%%%%%%%%%%%%%%%%%%%%%%%%%%%%%%%%%%%%%%%%%%%%%%%%%%%%%%%%%%%%%%%%%%%%%%
%despite the absence of looped chains and flower micelles.
%%%%%%%%%%%%%%%%%%%%%%%%%%%%%%%%%%%%%%%%%%%%%%%%%%%%%%%%%%%%%%%%%%%%%%%%%
It indicates that the transient network theory is useful tool 
not only for the study of rheological 
but also for thermodynamic properties of transient gels
when it is extended so that both the correlation among polymers 
and global structure of the network are taken into consideration.

This paper is organized as follows.
In section \ref{sectone2}, 
we will review assumptions and definitions employed in I.
They are also employed in this paper.
In section \ref{secSCdef}, linear viscoelasticities of the transient gel
will be studied within the framework of the Scanlan-Case criterion for active chains.
As the first step, only primary bridges are taken into consideration in this section.
Effect of superbridges will be discussed in section \ref{secsubri}.
In section \ref{resultsec}, liner rheology including effects of superbridges 
will be studied.
Section \ref{conlsec} will be devoted to a summary.

%%%%%%%%%%%%%%%%%%%%%%%%%%%%%%%%%%%%%%%%%%%%%%%%%%%%%%%%%%%%%%%%%%%%%%%%%%%%%
%%%%%%%%%%%%%%%%%%%%%%%%%%%%%%%%%%%%%%%%%%%%%%%%%%%%%%%%%%%%%%%%%%%%%%%%%%%%%
\section{Assumptions and Definitions of Fundamental Quantities}
\label{sectone2}

We consider athermal solutions of 
$n$ linear polymers (or primary chains) per unit volume.
Functional groups capable of forming junctions through noncovalent bonding
are locally embedded in both ends of the primary chain.
Common assumptions employed in this paper and in I are as follows:
1) any number of functional groups 
are allowed to be bound together to form one junction;
2) reactions among functional groups are allowed to occur only in a stepwise manner;
3) primary chains are Gaussian chains;
4) the Rouse relaxation time 
of the primary chain is much smaller than the characteristic time 
of the macroscopic deformation applied to the system
and the lifetime of association among functional groups;
5) the looped chain formed by a single primary chain is absent;
6) the molecular weight of the primary chain 
is much smaller than the entanglement molecular weight,
and hence effects of the topological interaction among chains is ignored.

The terminology used in this paper (and I) are as follows:
1) the number of functional groups forming a junction
(or the aggregation number), say $k$, 
is referred to as the junction multiplicity;
2) the junction with the multiplicity $k$ is called the $k$-junction;
3) the primary chain whose head is 
incorporated into the $k$-junction 
whereas whose tail is a member of the $\kp$-junction 
is referred to as the $(k,\kp)$-chain
(we virtually mark one end of each chain to 
identify a head and tail of the chain for convenience,
although physical properties of the chain is homogenous along the chain);
4) the primary chain whose one end,
irrespective of whether it is a head or tail,
is incorporated into the $k$-junction is called the $k$-chain.

As in I, we define $F_{k,\kp}(\br,t)d\br$
as the number of $(k,\kp)$-chains at time $t$
per unit volume with the head-to-tail vector $\br\sim\br+d\br$.
The total number of $(k,\kp)$-chains (per unit volume) is then given by 
\bea
\nu_{k,\kp}(t)=\int\!d\br~ F_{k,\kp}(\br,t)=\nu_{\kp,k}(t)
\eea
where two subscripts 
of $\nu_{k,\kp}(t)$ are exchangeable
due to the symmetry along the polymer chain.
The total number of primary chains 
\bea
n=\sum_{k\ge 1}\sum_{\kp\ge 1}\nu_{k,\kp}(t)
\label{nhozon33}
\eea
is conserved independent of time.
The number of $k$-chains is given by
\bea
\chi_{k}(t)=\sum_{\kp\ge1}\nu_{k,\kp}(t),
\label{defwateigi}
\eea
and then we can express the number of $k$-junctions as
\bea
\mu_k(t)=\frac{2\chi_k(t)}{k}.
\eea
The number of functional groups belonging to $k$-junctions is 
$k\mu_k(t)=2\chi_k(t)$ while the total number of functional groups is $2n$,
so that the probability that an arbitrary chosen functional group 
is in a $k$-junction can be expressed as
\bea
q_k(t)=\frac{\chi_k(t)}{n}.
\eea
Eq. (\ref{nhozon33}) is equivalent to the normalization condition
of $q_k$, i.e.,
\bea
\sum_{k\ge 1}q_k(t)=1.
\eea
The extent of reaction (or the probability for a 
functional group to be associated with other groups) is written as
\bea
\alpha(t)=\sum_{k\ge 2}q_k(t)=1-q_1(t).
\label{alphanama}
\eea

%%%%%%%%%%%%%%%%%%%%%%%%%%%%%%%%%%%%%%%%%%%%%%%%%%%%%%%%%%%%%%%%%%%%%%%%%%%%%
%%%%%%%%%%%%%%%%%%%%%%%%%%%%%%%%%%%%%%%%%%%%%%%%%%%%%%%%%%%%%%%%%%%%%%%%%%%%%
%%%%%%%%%%%%%%%%%%%%%%%%%%%%%%%%%%%%%%%%%%%%%%%%%%%%%%%%%%%%%%%%%%%%%%%%%%%%%

\section{Theory on the basis of the Scanlan-Case Criterion for Active Chains}
\label{secSCdef}

%%%%%%%%%%%%%%%%%%%%%%%%%%%%%%%%

\subsection{Global Structure of the Network}

Here, we briefly review how to treat the global structure of the
network according to refs. \cite{flory,graessley,tanaishi}.
Above a certain concentration of primary chains,
an infinite network (gel) is formed.
In the postgel regime, 
the extent of reaction $\alpha^{\prime}$ with regard to the functional groups
in the sol part 
is different from that $\alpha^{\prime\prime}$
in the gel part \cite{stocky12,flory}.
%%%%%%%%%%%%%%%%%%%%%%%%%%%%%%%%%%%%%%%%%%%%%%%%%%%%%%
%The theoretical argument given in this article is 
%based on Flory's picture \cite{flory}.
%%%%%%%%%%%%%%%%%%%%%%%%%%%%%%%%%%%%%%%%%%%%%%%%%%%%%%
Eq. (\ref{alphanama}) is the average extent of reaction for all functional groups
in the system, i.e., it can be expressed as
\bea
\alpha(t)=\alpha^{\prime}(t)w_S(t)+\alpha^{\prime\prime}(t)w_G(t)
\eea
where $w_S(t)$ is the fraction of the sol part and $w_G(t)=1-w_S(t)$ is 
that of the gel.
The sol fraction 
can be written as \cite{flory,graessley,tanaishi}
\bea
w_S(t)=\sum_{k\ge 1}q_k(t)\zeta_0^k,
\eea
where $\zeta_0$ is the probability that a randomly chosen unreacted group 
belongs to the sol part through its main chain.
In the pregel regime, we have only $\zeta_0=1$.
In the postgel regime, on the other hand,
we have $\zeta_0$ less than 1. 
Thus $\zeta_0$ is useful as an indicator of gelation.
%%%%%%%%%%%%%%%%%%%%%%%%%%%%%%%%%%%%%%%%%%%%%%%%%%%%%%%%%%%%%%%%%%%%%%%%%%%%%%%%%%
%We take account of the global structure of the system by making use of $\zeta_0$ 
%as some researcher did for equilibrium state \cite{flory,graessley,tanaishi}.
%%%%%%%%%%%%%%%%%%%%%%%%%%%%%%%%%%%%%%%%%%%%%%%%%%%%%%%%%%%%%%%%%%%%%%%%%%%%%%%%%%
A primary chain belongs to the sol
if its both ends are associated with the sol part,
so that the sol fraction can be also expressed as
\bea
w_S(t)=\Bigl(\sum_{k\ge 1}q_k(t)\zeta_0^{k-1}\Bigr)^2.
\eea
Therefore, $\zeta_0$ is the root of the following equation
\bea
x=u(x), \label{eqzeta0toku}
\eea
where 
\bea
u(x)\equiv\sum_{k\ge 1}q_k(t)x^{k-1}.
\label{unoteigi}
\eea
If (\ref{eqzeta0toku}) has more than one root, we must employ
the smallest one.

Now we consider the connectivity of a functional group to the gel network
according to the theoretical treatment by
Pearson and Graessley \cite{graessley}.
Let $\mu_{i,k}$ be the number of junctions with multiplicity $k$
that is connected to the gel network through $i$ paths ($0\le i \le k$).
Such a junction is called the ($i,k$)-junction in the following.
According to the multinomial theorem, it takes the form \cite{graessley, tanaishi}
\bea
& &\mu_{i,k}(t)=\mu_k(t)\frac{k!}{i!(k-i)!}\zeta_0^{k-i}(1-\zeta_0)^i.
\eea
Then the number of paths emersed from ($i,k$)-junction is
$\chi_{i,k}(t)=(i/2)\mu_{i,k}(t)$.
\begin{figure}[t]
\begin{center}
\includegraphics*[scale=0.5]{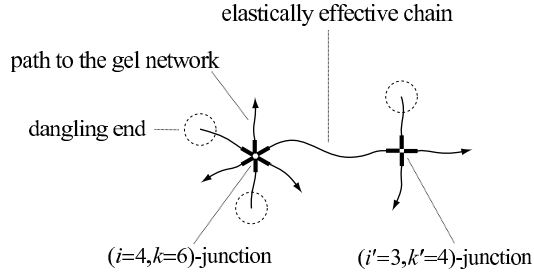}
\end{center}
\vspace*{-0.5cm}
\caption{
The classification of the junction by the multiplicity $k$
and the path connectivity $i$ to the gel network.
For example, a left-hand junction is formed by six functional groups
but it is connected with the gel network only through four paths 
(represented by arrows); the other two paths are connected with
dangling ends (indicated by dotted circles).
A primary chain whose both ends are connected to the junctions with $i\ge 3$
is elastically effective.
}
\label{isetsu}
\end{figure}
Here we employ the criterion of Scanlan \cite{scanlan} 
and Case \cite{case} to decide whether the primary chain
is active or not.
They suggested that the primary chain whose both ends are connected to
junctions with the path connectivity larger than or equal to 3 is
elastically effective (see Fig.\ref{isetsu}).
According to this criterion, the number of active chains
whose one end is belonging to $k$-junctions is 
\bea
\chi_k^{\eff}(t)
=\sum_{i\ge 3}^k\chi_{i,k}(t)
=\chi_k(t)(1-\zeta_0)
\bigl[
1-\zeta_0^{k-1}-(k-1)\zeta_0^{k-2}(1-\zeta_0)
\bigr].
\eea
Note that $\chi_1^{\eff}(t)=\chi_2^{\eff}(t)=0$.
The total number of active chains is then
given by \cite{tanaishi}
\bea
\nu^{\eff}(t)=\sum_{k\ge3}\chi_k^{\eff}(t)
=n(1-\zeta_0)^2\bigl(1-u^{\prime}(\zeta_0)\bigr), \label{tanisaeff}
\eea
where we have used a relation $\zeta_0=u(\zeta_0)$.
The number of active $(k,\kp)$-chains
(i.e., the chain whose one end is connected to the 
($i,k$)-junction with $i\ge 3$
and the other end is belonging to the $(i^{\prime},\kp)$-junction
with $i^{\prime}\ge 3$) can be defined as
\bea
\nu^{\eff}_{k,\kp}(t)
=\frac{\chi_k^{\eff}(t)\chi_{\kp}^{\eff}(t)}{\nu^{\eff}(t)}.
\label{neffkkp}
\eea
The following relation
\bea
\sum_{k,\kp\ge 3}\nu^{\eff}_{k,\kp}(t)=\nu^{\eff}(t)
\eea
holds as it should be.

%%%%%%%%%%%%%%%%%%%%%%%%%%%%%%%%%%%%%%%
%%%%%%%%%%%%%%%%%%%%%%%%%%%%%%%%%%%%%%%
\subsection{Time-Development of Chains}

The number of ($k,\kp$)-chains with the head-to-tail vector $\br$
obeys the following time-evolution equation:
\bea
\frac{\partial F_{k,\kp}(\br,t)}{\partial t}+\nabla\!\cdot\!
\Bigl(\dot{\br}_{k,\kp}(\br,t) F_{k,\kp}(\br,t)\Bigr)
=W_{k,\kp}(\br,t)
~~(\mbox{for}~k,\kp\ge 3),
\label{baseeqige3}
\eea
where $\dot{\br}_{k,\kp}(\br,t)$ is the rate of deformation 
of the head-to-tail vector $\br$ of the ($k,\kp$)-chain. 
When a macroscopic deformation is applied to the gel,
only active chains deform.
%%%%%%%%%%%%%%%%%%%%%%%%%%%%%%%%%%%%%%%%%%%%%%%%%%%%%
% これは書かない．
%elastically ineffective chains
%are effectively in equilibrium state 
%and do not deform.
%%%%%%%%%%%%%%%%%%%%%%%%%%%%%%%%%%%%%%%%%%%%%%%%%%%%%%%%%%%%%%
%  下を加えるべきか悩む．
%%%%%%%%%%%%%%%%%%%%%%%%%%%%%%%%%%%%%%%%%%%%%%%%%%%%%%%%%%%%%%%%%%%%%%%%%%%%%%%%%%%%%%
%\footnote{dangling endが
%大きいときは，幾何学的な相互作用によって変化するかも．TIの論文に書いてあるように．}.
%%%%%%%%%%%%%%%%%%%%%%%%%%%%%%%%%%%%%%%%%%%%%%%%%%%%%%%%%%%%%%%%%%%%%%%%%%%%%%%%%%%%%%
%
Some ($k,\kp$)-chains are active but the other
($k,\kp$)-chains are not because each junction has the different path connectivity
even if it has the same multiplicity.
To take this into account, we put
\bea
\dot{\br}_{k,\kp}(t)=P_{k,\kp}(t)\hat{\kappa}(t)\br
\label{nonafifine2}
\eea
where
$\hat{\kappa}(t)$ is the rate of deformation tensor applied to the gel,
and
\bea
P_{k,\kp}(t)\equiv\frac{\nu^{\eff}_{k,\kp}(t)}{\nu_{k,\kp}(t)}
\label{Pkakuteig}
\eea
is the probability for a ($k,\kp$)-chain to be active.
Eq. (\ref{nonafifine2}) states that active chains deform affinely to the 
macroscopic deformation {\it on average}.
Eq. (\ref{baseeqige3}) holds only for $k,\kp\ge3$ because
both 1-junction and 2-junction cannot have path connectivity more than
or equal to 3 and chains connected with such junctions do not deform.
We assume that these elastically ineffective primary chains are Gaussian chains;
i.e., the probability distribution function that they take the head-to-tail 
vector $\br$ is
\bea
f_0(\br)\equiv\left(\frac{3}{2\pi Na^2}\right)^{3/2}
\exp\left(-\frac{3|\br|^2}{2Na^2}\right)
\eea
where $N$ and $a$ is the number and the length of the repeat unit
constructing a primary chain, respectively. 
For example, the distribution function for the dangling chain is written as
$F_{k,1}(\br,t)=\nu_{k,1}(t)f_0(\br)$.
The right-hand side of (\ref{baseeqige3}) represents the net increment of 
the ($k,\kp$)-chain per unit time due to the reaction
between the end groups on the ($k,\kp$)-chain
and the groups on the other chains.
As shown in I, it takes the form
\bea
W_{k,\kp}(\br,t)=
& &-\bigl[
\beta_k(r)+\beta_{\kp}(r)
+B_k(t)+B_{\kp}(t)+P_k(t)+P_{\kp}(t)
\bigr]F_{k,\kp}(\br,t) \non\\
& &+\bigl[
p_{k-1}(t)\nu_{1,\kp}(t)+p_{\kp-1}(t)\nu_{k,1}(t)
\bigr]f_0(\br) \non\\
& &+B_{k+1}(t)F_{k+1,\kp}(\br,t)+B_{\kp+1}(t)F_{k,\kp+1}(\br,t) \non\\
& &+P_{k-1}(t)F_{k-1,\kp}(\br,t)+P_{\kp-1}(t)F_{k,\kp-1}(\br,t)
~~~~(\mbox{for}~k,\kp\ge 3)
\label{hannouactive}
\eea
where $\beta_k(r)$ is the probability
that a functional group is dissociated
from the $k$-junction per unit time,
and $p_k(t)$ is the probability for
an unreacted group to be connected with the $k$-junction per unit time.
As in I, we assume that the dissociation rate 
does not depend on the junction multiplicity
nor the end-to-end length of the primary chain 
that is connected to the junction ($\beta_k(r)=\beta$).
On the other hand, the connection rate $p_k$ of the functional group
with a $k$-junction is assumed to be proportional to the volume fraction of
the $k$-junction, and we put $p_k(t)=\beta \lambda\psi q_k(t)h_k$
where $\lambda=\exp(\epsilon/k_BT)$ is the association constant
($\epsilon$ is the binding energy for the attraction of functional groups),
$\psi=2nv_0$ is the volume fraction of the functional group ($v_0$ is the effective 
volume of a segment), and $h_k$ is the factor depending on $k$ 
that give a limitation on the junction multiplicity.
In the following, we use $c\equiv \lambda \psi=2\lambda \phi/N$ 
as a reduced polymer concentration ($\phi\equiv Nnv_0$ is the volume fraction of 
the primary chain).
Under these assumptions, $B_k(t)$ and $P_k(t)$ in (\ref{hannouactive}) 
are expressed as $B_k=\beta(k-1)$ and $P_k(t)=\beta kcq_1(t)h_k$, respectively.

The kinetic equation for ($k,\kp$)-chains 
has the same form as the one derived in I, i.e.,
\bea
\frac{d\nu_{k,\kp}(t)}{dt}=w_{k,\kp}(t)+w_{\kp,k}(t)
~~~~~(\mbox{for} ~k,\kp\ge 1),
\label{numefqw}
\eea
where
\
\begin{subequations}\label{www2}
\bea
& &w_{k,\kp}(t)=-\beta k (1+cq_1(t)h_k)\nu_{k,\kp}(t)
+\beta k \nu_{k+1,\kp}(t)
+(k-1)\beta cq_1(t) h_{k-1} \nu_{k-1,\kp}(t) \non\\
& &\hspace*{1.9cm}+\beta c h_{k-1}q_{k-1}(t)\nu_{1,\kp}(t)
~~~~~(\mbox{for} ~k\ge 2), \label{wnodifk1}\\
& &w_{1,\kp}(t)=\beta\Bigl(\sum_{l\ge 2}\nu_{l,\kp}(t)+\nu_{2,\kp}(t)\Bigr)
-\beta c \Bigl(\sum_{l\ge 1}h_lq_l(t)+h_1q_1(t)\Bigr)\nu_{1,\kp}(t).
\eea
\end{subequations}
Note that (\ref{baseeqige3}) reduces to (\ref{numefqw}) 
by integrating it with respect to $\br$ (for $k,\kp\ge 3$).
The kinetic equation for $k$-chains is obtained by 
taking a summation in (\ref{numefqw}) for overall $\kp$.
It takes the form
\bea
\frac{dq_{k}(t)}{dt}=\tilde{v_k}(t)
~~~~~(\mbox{for} ~k\ge 1), \label{utiele}
\eea
where 
\begin{subequations}\label{kinechi200mato}
\bea
\tilde{v}_k(t)
=& &-\beta k\bigl(q_k(t)-q_{k+1}(t)\bigr) 
+\beta ck\bigl(h_{k-1}q_{k-1}(t)-h_kq_k(t)\bigr)q_1(t) 
~~~(\mbox{for}~k\ge 2),~~~~~\label{kinechi200}\\
\tilde{v}_1(t)
=& &\beta\Bigl(\sum_{l\ge 2}q_{l}(t)+q_2(t)\Bigr)
-\beta c\Bigl(\sum_{l\ge 1}h_lq_l(t)+h_1q_1(t)\Bigr)q_1(t).
\label{kinechi100}
\eea
\end{subequations}
Once $q_k(t)$ is derived by solving (\ref{utiele}), we can obtain 
the number of ($k,\kp$)-chains from the relation
$\nu_{k,\kp}(t)=nq_k(t)q_{\kp}(t)$ and $\zeta_0$ 
from (\ref{eqzeta0toku}), respectively.
Then $\nu_{k,\kp}^{\eff}(t)$ 
and $P_{k,\kp}(t)=\nu_{k,\kp}^{\eff}(t)/\nu_{k,\kp}(t)$ can be obtained.
By putting $P_{k,\kp}(t)$ into (\ref{baseeqige3}) and solving the equations,
we can derive $F_{k,\kp}$.

We study two special models of junctions 
by putting a limitation on the multiplicity,
i.e., the saturating junction model 
and the fixed multiplicity model \cite{tanastock,tanaishi,in1}.
In the saturating junction model, we allow junctions
to take only a limited range of multiplicity $k=1,2,...s_m$.
This condition is realized 
by putting $h_k=1$ for $1\le k \le s_m-1$ and $h_k=0$ for $k\ge s_m$. 
If $s_m=\infty$, junction can take any value of multiplicity without limitation.
In the fixed multiplicity model,
all junctions can take only the same multiplicity $s$ (except for $k=1$).
This situation can be approximately attained by introducing 
a small quantity $\delta(\ll1)$ and by putting
$h_k=\delta$ for $1\le k< s-1$, $h_{s-1}=\delta^{-(s-2)}$, and
$h_k=0$ for $k>s-1$.\cite{in1} We set $\delta=0.01$ in most cases.

%%%%%%%%%%%%%%%%%%%%%%%%%%%%%%%%%%%%%%%%%%%%%%%%%%%%%%%%%%%%%%%%%%%%%%%%%%%%%%%%%%%%%%%
%%%%%%%%%%%%%%%%%%%%%%%%%%%%%%%%%%%%%%%%%%%%%%%%%%%%%%%%%%%%%%%%%%%%%%%%%%%%%%%%%%%%%%%

\subsection{Equilibrium Properties}
\label{subseceqi}

By putting $dq_k/dt=0$ in (\ref{utiele}), we can obtain $q_k$ in equilibrium state.
It turns out to be
\bea
q_k=\gamma_kc^{k-1}q_1^k
\label{eqcon}
\eea
where $\gamma_k=\prod_{l=1}^{k-1}h_l$ for $k\ge 2$ and $\gamma_1=1$.
The fraction of unreacted 
groups $q_1$ is determined from 
the normalization condition $\gamma(z)q_1=1$,
where $\gamma(z)\equiv\sum_{k\ge 1}\gamma_kz^{k-1}$ and $z\equiv cq_1$.
The function defined by (\ref{unoteigi}) can be expressed
as $u(x)=\gamma(zx)/\gamma(z)$ in the equilibrium state,
and hence $\zeta_0$ is a solution of 
the following equation for a given $z$ (or $c$):
\bea
x=\frac{\gamma(xz)}{\gamma(z)}.
\label{zeta0teij}
\eea

\begin{figure}[t]
\begin{center}
\includegraphics*[scale=0.5]{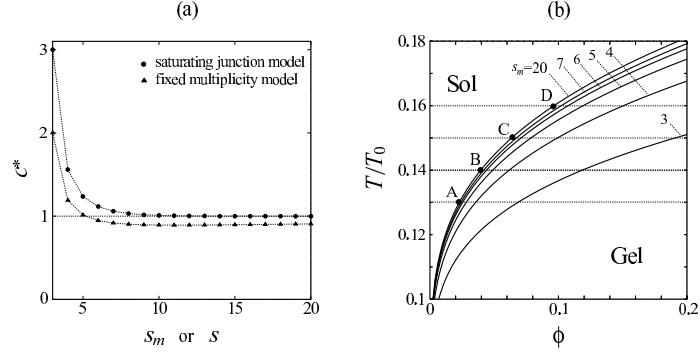}
\end{center}
\vspace*{-0.5cm}
\caption{(a) Reduced sol/gel transition concentration
for the saturating junction model (circles) 
plotted against the maximum multiplicity $s_m$ 
and for fixed multiplicity model (triangles)
plotted against the multiplicity $s$. 
(b) Sol/gel transition curves drawn in the temperature-volume fraction plane
for saturating junction model with $\lambda_0=1$ and $N=100$.
The maximum multiplicity is varying from curve to curve.
See also Fig.\ref{satgeta-T}.}
\label{solge}
\end{figure}

In the case that the junction can take any value of 
multiplicity without limitation, for example,
$z$ is smaller than 1 (see I) and hence $\gamma(z)=1/(1-z)$ 
(we can put $h_k=1$ for all $k$ in this case).
Thus $\zeta_0$ is obtained as the smallest root of 
the equation $x=(1-z)/(1-xz)$, i.e.,
\bea
\zeta_0= \left\{
\begin{array}{@{\,}ll}
1     &  (0\le z<z^*) \\
1/z-1 & (z^*<z\le1) 
\end{array}
\right.,
\label{gasmmateigisd}
\eea
where $z^*=1/2$.
Note that $z^*$ is interpreted as the 
critical value of the parameter $z$ for the sol/gel transition.
The number of elastically effective chain given by (\ref{tanisaeff}) 
is then analytically expressed as a function of $z$:
\bea
\nu^{\eff}
&=&\left\{
\begin{array}{@{\,}ll}
0     &  (0\le z<z^*) \\
n\bigl(2(z-z^*)/z\bigr)^3& (z^*<z\le1) 
\end{array}
\right.,
\eea
We can also express these quantities as a function of $c$ 
by the use of the relation $z=c/(1+c)$, that is,
\bea
\zeta_0= \left\{
\begin{array}{@{\,}ll}
1     &  (c<c^*) \\
1/c & (c>c^*) 
\end{array}
\right.,
\label{gasmmateigisd2}
\eea
and
\bea
\nu^{\eff}&=&
\left\{
\begin{array}{@{\,}ll}
0     &  (c<c^*) \\
n\bigl((c-c^*)/c\bigr)^3& (c>c^*) 
\end{array}
\right.,
\eea
where $c^*=1$ is the critical reduced concentration for gelation
(see Fig.\ref{solge} (a) for the case of limited multiplicity).
Near the sol/gel transition point,
the number of elastically effective chains 
increases as the cube of the concentration deviation,\cite{tanaishi} i.e.,
$\nu^{\eff}\simeq \Delta^3$ where $\Delta\equiv (c-c^*)/c^*$.
In the high concentration limit, on the other hand, 
all primary chains become elastically effective, i.e.,
$\nu^{\eff}\to n$ for $c\to\infty$.

In general, the sol/gel transition point is obtained as the point
at which $\zeta_0$ becomes lower than 1.
This is equivalent to the point where the weight-average molecular weight 
of the cluster diverges.
For polycondensation by multiple reaction,
Fukui and Yamabe have shown that this condition
(an appearance of the macroscopic cluster) 
is given by\cite{fukuyama} 
\bea
(f_w-1)(\mu_w-1)=1 \label{hukuyamasiki}
\eea
where $f_w$(=2 for the present model) 
is the weight-average functionality of the primary chain,
and $\mu_w$
is the weight-average multiplicity of the junctions given by
\bea
\mu_w\equiv \sum_{k\ge 1}kq_k=1+\frac{z\gamma^{\prime}(z)}{\gamma(z)}.
\eea

A boundary curve separating the sol region and gel region in
the temperature-concentration plane can be drawn by the use of the equation
$2\lambda(T^*) \phi^*/N=c^*$ where $T^*$ and $\phi^*$ are 
the critical temperature and volume fraction of the primary chain, respectively,
and $c^*$ is obtained according to the procedure stated above.
Since the binding free energy is comprised of the energy part 
$\epsilon_0$ and the entropy part $S_0$,
the association constant is rewritten as
$\lambda(T)=\lambda_0\exp(T_0/T)$ with $\lambda_0\equiv \exp(-S_0/k_B)$ and 
$T_0\equiv \epsilon_0/k_B$.
Thus we have
\bea
T^*=T_0/\log\left(Nc^*/2\lambda_0\phi^*\right).
\label{tstarkime}
\eea
Fig.\ref{solge} (b) shows the sol/gel boundary lines given by (\ref{tstarkime}) 
for the saturating junction model as an example.

%%%%%%%%%%%%%%%%%%%%%%%%%%%%%%%%%%%%%%%%%%%%%%%%%%%%%%%%%%%%%%%%%%%%%%%%%%%%%
%%%%%%%%%%%%%%%%%%%%%%%%%%%%%%%%%%%%%%%%%%%%%%%%%%%%%%%%%%%%%%%%%%%%%%%%%%%%%
%%%%%%%%%%%%%%%%%%%%%%%%%%%%%%%%%%%%%%%%%%%%%%%%%%%%%%%%%%%%%%%%%%%%%%%%%%%%%
%%%%%%%%%%%%%%%%%%%%%%%%%%%%%%%%%%%%%%%%%%%%%%%%%%%%%%%%%%%%%%%%%%%%%%%%%%%%%
\subsection{Dynamic-Mechanical and Viscoelastic Properties}

We now apply a small oscillatory shear deformation 
with an amplitude $\epsilon$ to the gel network.
A $xy$ component of the rate of this deformation tensor 
is represented by $\kappa_{xy}(t)=\epsilon\omega\cos \omega t$ 
while the other components are 0 
($\omega$ is the frequency of the oscillation).
Let us expand $F_{k,\kp}(\br,t)$ 
with respect to the powers of $\epsilon$ up to the first order:
\bea
F_{k,\kp}(\br,t)=F^{(0)}_{k,\kp}(\br)+\epsilon F^{(1)}_{k,\kp}(\br,t).
\label{f1teigi30002}
\eea
We can put (see I)
\begin{subequations}\label{Gkktotyuu300}
\bea
& &F^{(0)}_{k,\kp}(\br)=\nu_{k,\kp}f_0(\br), \label{f1teigi3000}\\
& &F^{(1)}_{k,\kp}(\br,t)=
\Bigl(
g^{\prime}_{k,\kp}(\omega)\sin\omega t+g^{\prime\prime}_{k,\kp}(\omega)\cos\omega t
\Bigl)
\frac{3xy}{Na^2}f_0(\br).
\label{f1teigi3}
\eea
\end{subequations}
The number of ($k,\kp$)-chains does not depend on time as far as 
the small shear deformation is concerned,\cite{tanaed1,tanaed2,intana1}
so that $\nu_{k,\kp}(t)$ and $q_{k}(t)$ can be represented by 
their equilibrium values $\nu_{k,\kp}$ and $q_{k}$ given above, respectively. 
Then the probability for a ($k,\kp$)-chain to be elastically effective is given by
its equilibrium value
$P_{k,\kp}=\nu^{\eff}_{k,\kp}/\nu_{k,\kp}$ with
\bea
\nu_{k,\kp}^{\eff}=\nu_{k,\kp}
\frac{\bigl[1-\zeta_0^{k-1}-(k-1)\zeta_0^{k-2}(1-\zeta_0)\bigr]
\bigl[1-\zeta_0^{\kp-1}-(\kp-1)\zeta_0^{\kp-2}(1-\zeta_0)\bigr]}
{1-z\gamma^{\prime}(\zeta_0z)/\gamma(z)}.
\hspace*{1.5cm}
\label{neureff}
\eea
The in-phase $g^{\prime}_{k,\kp}(\omega)$ 
and out-of-phase $g^{\prime\prime}_{k,\kp}(\omega)$
amplitude of $F^{(1)}_{k,\kp}$ are directly related with 
the storage and loss modulus of
$(k,\kp)$-chains through the relations
$G^{\prime}_{k.\kp}(\omega)=k_BTg^{\prime}_{k,\kp}(\omega)$
and $G^{\prime\prime}_{k.\kp}(\omega)=k_BTg^{\prime\prime}_{k,\kp}(\omega)$,
respectively. 
Since primary chains are symmetric along the chain, 
two subscripts are exchangeable, i.e.,
$g_{k,\kp}^{\prime}=g_{\kp,k}^{\prime}$ 
(and $g_{k,\kp}^{\prime\prime}=g_{\kp,k}^{\prime\prime}$). 
Note that $g_{k,1}^{\prime}=g_{k,1}^{\prime\prime}=
g_{k,2}^{\prime}=g_{k,2}^{\prime\prime}= 0$ for $k\ge 1$
since 2-chains and 1-chains 
are effectively in equilibrium state. 
By substituting (\ref{f1teigi30002}) 
with (\ref{Gkktotyuu300}) into (\ref{baseeqige3}),
we have a set of equations for 
$g^{\prime}_{k,\kp}$ and $g^{\prime\prime}_{k,\kp}$ as follows
\begin{subequations}\label{kretoek2}
\bea
& &g^{\prime}_{k,\kp}=
\left(-Q_{k,\kp}g^{\prime\prime}_{k,\kp}
+B_{k+1}g^{\prime\prime}_{k+1,\kp}
+B_{\kp+1}g^{\prime\prime}_{k,\kp+1}
+P_{k-1}g^{\prime\prime}_{k-1,\kp}
+P_{\kp-1}g^{\prime\prime}_{k,\kp-1}\right)/\omega+\nu_{k,\kp}^{\eff}, 
\label{gpkores}
~~~~~~~~~~~~\\
& &g^{\prime\prime}_{k,\kp}
=\left(Q_{k,\kp}g^{\prime}_{k,\kp}
-B_{k+1}g^{\prime}_{k+1,\kp}
-B_{\kp+1}g^{\prime}_{k,\kp+1}
-P_{k-1}g^{\prime}_{k-1,\kp}
-P_{\kp-1}g^{\prime}_{k,\kp-1}\right)/\omega \\
& &\hspace*{13cm}(\mbox{for~}k,\kp\ge 3) \non
\eea
\end{subequations}
where 
$B_k=\beta(k-1), P_k=\beta k cq_1h_k$,
$Q_{k,\kp}(t)\equiv\beta k(1+cq_1h_k) + \beta\kp(1+cq_1h_{\kp})$, 
and $\nu^{\eff}_{k,\kp}$ is given by (\ref{neureff}).
It should be emphasized here that the last term in the right-hand side of 
(\ref{gpkores}) is $\nu^{\eff}_{k,\kp}$ instead of $\nu_{k,\kp}$ 
(see I as a reference).
This is a consequence of the assumption (\ref{nonafifine2}).

\begin{figure}[t]
\begin{center}
\includegraphics*[scale=0.5]{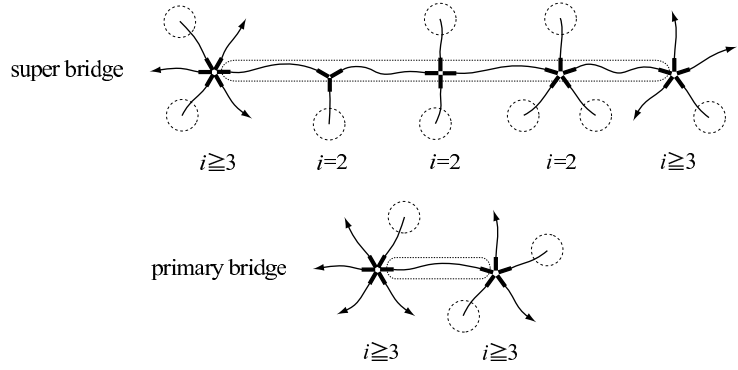}
\end{center}
\vspace*{-0.5cm}
\caption{Upper: schematic representation of a superbridge 
(surrounded by dotted ellipsoid).
Lower: schematic representation of a primary bridge 
(or active primary chain).
}
\label{super2}
\end{figure}

The total moduli within the framework of the Scanlan-Case criterion 
for elastically effective chains are given 
by taking a summation of $G^{\prime}_{k.\kp}(\omega)$ and 
$G^{\prime\prime}_{k.\kp}(\omega)$ over $k,\kp\ge 3$, i.e.,
\begin{subequations}\label{koreGpGPP}
\bea
& &G^{\prime}(\omega)=
k_BT\sum_{k\ge 3}\sum_{\kp\ge 3}g^{\prime}_{k,\kp}(\omega), \\
& &G^{\prime\prime}(\omega)=
k_BT\sum_{k\ge 3}\sum_{\kp\ge 3}g^{\prime\prime}_{k,\kp}(\omega).
\label{gppllep}
\eea
\end{subequations}
They are well described in terms of the Maxwell model 
with a single relaxation time (not shown here).
In the high frequency limit, (\ref{gpkores}) becomes
$g^{\prime}_{k,\kp}(\omega\to\infty)=\nu_{k,\kp}^{\eff}$.
Therefore, the plateau modulus 
defined by $G_{\infty}\equiv G^{\prime}(\omega\to\infty)$
can be expressed as
\bea
G_{\infty}
=\nu^{\eff}k_BT
=nk_BT(1-\zeta_0)^2\left(1-\frac{z\gamma^{\prime}(\zeta_0z)}{\gamma(z)}\right).
\label{tanaishin}
\eea
The reduced plateau modulus $G_{\infty}/(nk_BT)$
coincides with the number of active chains
derived by Tanaka and Ishida
for telechelic polymers.\cite{tanaishi}
%
%%%%%%%%%%%%%%%%%%%%%%%%%%%%%%%%%%%%%%%%%%%%%%
%  HEURの略記について，後でちゃんと書かねば．%
%%%%%%%%%%%%%%%%%%%%%%%%%%%%%%%%%%%%%%%%%%%%%%
%
As they have shown, 
it agrees well with
experimental data for aqueous solution of 
hydrophobically modified ethylene oxide-urethane copolymers (called HEUR)
reported by Annable {\it et al.}.\cite{annable1}
On the other hand, the relaxation time $\tau$ of the gel determined from 
the peak position of (\ref{gppllep})
does not agree well with experiments as in the case of I
because it depends on $c$ only weakly.
This discrepancy can be ascribed to the absence of 
{\it superbridges} in elastically effective chains.
A superbridge is a linear cluster of primary chains
whose backbone includes a number of junctions with 
the path connectivity $i=2$ to the gel network.
Both ends of a superbridge are connected to junctions 
with the path connectivity $i\ge 3$.
Fig.\ref{super2} shows an example of the superbridge
whose backbone is formed by four primary chains.
In the Scanlan-Case criterion, superbridges are not regarded as
the elastically effective chains;
only {\it primary bridges}, or active primary chains (see Fig.\ref{super2}),
are assumed to be responsible for the elasticity of the network.
If we take superbridges into account,
not only that the number of active chains becomes larger,
but that the relaxation time of the network becomes shorter since
their lifetime is shorter than that of primary bridge due to breakable
node inside the backbone.
We will discuss the effects of superbridges on $G^{\prime}(\omega)$ 
and $G^{\prime\prime}(\omega)$
in the next section.\footnote{We use the nomenclature 
superbridge after ref.\cite{annable1} 
where a cluster formed by connecting
flower micelles linearly through bridges is called the superbridge.
Their effects on the relaxation time is roughly estimated in this reference.
In this paper, we count the number of superbridges
in more detailed way by making use of the path connectivity
to study their effect on the relaxation time.}

%%%%%%%%%%%%%%%%%%%%%%%%%%%%%%%%%%%%%%%%%%%%%%%%%%%%%%%%%%%%%%%%%%%%%%%%%%%%%
%%%%%%%%%%%%%%%%%%%%%%%%%%%%%%%%%%%%%%%%%%%%%%%%%%%%%%%%%%%%%%%%%%%%%%%%%%%%%
%%%%%%%%%%%%%%%%%%%%%%%%%%%%%%%%%%%%%%%%%%%%%%%%%%%%%%%%%%%%%%%%%%%%%%%%%%%%

\section{Effects of Superbridges}
\label{secsubri}

\subsection{Number of Superbridges}
\label{subsecsubri}

Let $m(i,\ip)$ be the number of primary chains whose one end is connected to the 
junction with the path connectivity $i$ to the gel 
whereas the other end is 
connected to the junction with the connectivity $\ip$.
The number of primary bridges is then represented as\footnote{We
denote
the number of primary bridge as $\nu^{\eff}_{SC}$ instead of $\nu^{\eff}$
in the rest of this article.}
\bea
m(i\ge 3,\ip\ge 3)
=n(1-\zeta_0)^2\left(1-\frac{z\gamma^{\prime}(\zeta_0z)}{\gamma(z)}\right)
\equiv\nu^{\eff}_{SC}.
\eea
The total number of primary chains 
incorporated into the gel through both ends
is
\bea
& &m(i\ge 2,\ip\ge 2)=\sum_{k\ge2}\sum_{i\ge 2}\chi_{i,k}=n(1-\zeta_0)^2
\equiv \tilde{\nu}^{\eff},
\eea
and
the number of primary chains 
comprising backbones of 
superbridges is
\bea
& &m(i\ge 2,\ip= 2)=\sum_{k\ge 2}\chi_{2,k}
=n(1-\zeta_0)^2\frac{z\gamma^{\prime}(\zeta_0z)}{\gamma(z)}
\equiv \nu^{\eff}_{pseud}.
\eea
A relation $\tilde{\nu}^{\eff}=\nu^{\eff}_{SC}+\nu^{\eff}_{pseud}$
holds as it should be.
Near the sol/gel transition concentration (or $\Delta\!=\!(c-c^*)/c^*\!\ll\!1$), 
these quantities increase as
$\nu_{pseud}^{\eff}\sim\tilde{\nu}^{\eff}\sim\Delta^2$ 
(and $\nu_{SC}^{\eff}\sim\Delta^3$)
because $1-\zeta_0$ is proportional to $\Delta$
while $z\gamma^{\prime}(\zeta_0z)/\gamma(z)$ is proportional to $c$.
\begin{figure}[t]
\begin{center}
\includegraphics*[scale=0.5]{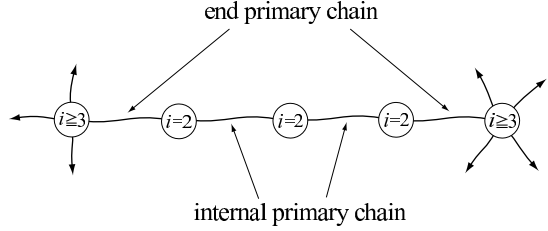}
\end{center}
\vspace*{-0.5cm}
\caption{A schematic of a superbridge (only paths to the network are drawn).
It is comprised of 
two end primary chains and several internal primary chains.
The number of end primary chains is written as $m(i\ge 3,\ip=2)$
whereas the number of internal primary chains is $m(i= 2,\ip=2)$.
}
\label{super7}
\end{figure}
Since the superbridge is comprised of 
two end primary chains and a number of internal primary chains
(see Fig.\ref{super7}), we have a relation
$\nu^{\eff}_{pseud}=m(i=2,\ip=2)+m(i\ge3,\ip=2)$ where
\bea
& &m(i=2,\ip=2)=
\sum_{k\ge 2}\sum_{\kp\ge 2}
\tilde{\nu}^{\eff}
\frac{\chi_{2,k}}
{
\tilde{\nu}^{\eff}
}
\frac{\chi_{2,\kp}}
{
\tilde{\nu}^{\eff}
}
=\frac{\bigl(\nu^{\eff}_{pseud}\bigr)^2}
{
\tilde{\nu}^{\eff}
} \label{k2jiu}
\eea
is the number of internal primary chains, and
\bea
& &m(i\ge3,\ip=2)=
\nu^{\eff}_{pseud}-m(i=2,\ip=2)
=\frac{\nu^{\eff}_{SC}\cdot\nu^{\eff}_{pseud}}
{\tilde{\nu}^{\eff}}
\eea
is the number of end primary chains.
In (\ref{k2jiu}), $\chi_{2,k}/\tilde{\nu}^{\eff}$
is the probability for a chain in the network to be connected with 
the ($i\!=\!2,k$)-junction, and hence
$\tilde{\nu}^{\eff}(\chi_{2,k}/\tilde{\nu}^{\eff})(\chi_{2,\kp}/\tilde{\nu}^{\eff})$
is the number of primary chains whose one end is connected with
the ($i\!=\!2,k$)-junction while the other end
is belonging to the ($\ip\!=\!2,\kp$)-junction.
The number of superbridges is a half of the number of
end primary chains of superbridges, i.e.,
\bea
\nu^{super}=\frac{1}{2}m(i\ge 3,\ip=2)
=\frac{\nu^{\eff}_{SC}\cdot\nu^{\eff}_{pseud}}
{2\tilde{\nu}^{\eff}}.
\eea
Therefore, the total number of elastically effective chain 
is turned out to be
\bea
\nu^{\eff}_{total}&=&\nu^{\eff}_{SC}+\nu^{super} \non\\
&=&n(1-\zeta_0)^2\left(1-\frac{z\gamma^{\prime}(\zeta_0z)}{\gamma(z)}\right)
\left(1+ 
\frac{z\gamma^{\prime}(\zeta_0z)}{2\gamma(z)}
\right).
\eea
\begin{figure}[t]
\begin{center}
\includegraphics*[scale=0.5]{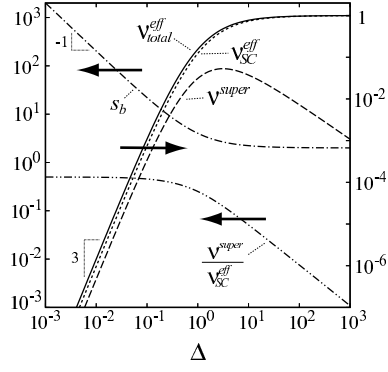}
\end{center}
\vspace*{-0.5cm}
\caption{
The total number of elastically effective chain (solid line),
the number of primary bridges
or elastically effective chains defined on the basis of the Scanlan-Case criterion 
(dotted line), the number of super bridge (broken line),
the number of primary chains 
per a super bridge (dash-dotted line),
the ratio between the number of superbridges and 
the number of primary bridges (dash-double dotted line)
plotted against relative concentration deviation $\Delta=(c-c^*)/c^*$
for the saturating junction model with the maximum multiplicity fixed at $s_m=15$.
}
\label{superfig}
\end{figure}
Fig.\ref{superfig} shows $\nu^{\eff}_{total}$ together with 
the number $\nu^{\eff}_{SC}$ of primary bridges, 
the number $\nu^{super}$ of superbridges, 
and the relative amount of the superbridges 
$\nu^{super}/\nu^{\eff}_{SC}$ compared with primary bridges
as a function of $\Delta$ for the saturating junction model ($s_m\!=\!15$).
The number of superbridges increases ($\nu^{super}\sim\Delta^3$)
near the sol/gel transition concentration,
but it decreases at higher concentration since 
the number of dangling ends decreases.
Thus a peak appears in $\nu^{super}$ at a modest concentration. 
It should be emphasized that $\nu^{super}/\nu^{\eff}_{SC}$
increases with decreasing $\Delta$, 
and it finally reaches 0.5 for $\Delta\to0$
although both $\nu^{super}$ and $\nu^{\eff}_{SC}$ become close to 0 in this limit.
It indicates that the effect of superbridges cannot be ignored 
as compared with primary bridges especially 
in the vicinity of the sol/gel transition point.
Fig.\ref{superfig} also show the number 
of primary chains forming a superbridge given by
\bea
s_b\equiv \frac{\nu^{\eff}_{pseud}}{\nu^{super}}.
\label{sbdef2}
\eea
With decreasing $\Delta$,
many primary chains become incorporated into
superbridges as $s_b\sim 1/\Delta$ (for $\Delta\ll1$)
indicating that the superbridge becomes longer along the backbone.
With increasing concentration,
on the contrary, 
$s_b$ approaches to 2 since $m(i=2,\ip=2)$ becomes close to 0.
Summarizing, in the vicinity of the sol/gel transition concentration,
(i) the superbridge is comparable in number to the primary bridge
although both are few, and
(ii) the superbridge is infinitely long.

%%%%%%%%%%%%%%%%%%%%%%%%%%%%%%%%%%%%%%%%%%%%%%%%%%%%%%%%%%%%%%%
\subsection{Breakage Rate of Superbridge}

\begin{figure}[t]
\begin{center}
\includegraphics*[scale=0.6]{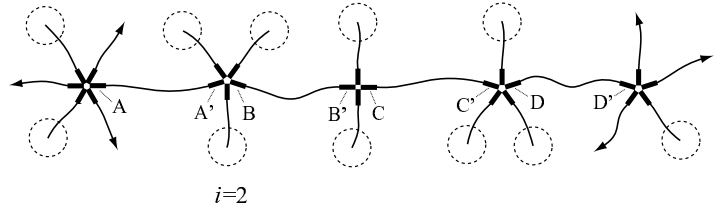}
\end{center}
\vspace*{-0.5cm}
\caption{An example of the superbridge whose backbone is 
comprised of $s_b$(=4) primary chains.
%If one of $2s_b(=8)$ functional groups A,A',...,D,D' is disconnected from 
%the junction, it breaks.
}
\label{super8}
\end{figure}

Let us here focus on a primary chain whose one end
is connected to the junction (say A)
with the path connectivity $i_{\mbox{\scriptsize{A}}}\ge3$
while the other end is belonging to the junction (A') 
with the path connectivity
$i_{\mbox{\scriptsize{A'}}}\ge2$.
%as discussed above.
Such a primary chain is elastically effective.\footnote{
Internal primary chains of the superbridge
(i.e., chains whose both ends are connected to the junctions
with the path connectivity two)
%$i_{\mbox{\tiny{A}}}=i_{\mbox{\tiny{A'}}}=2$)
are also elastically effective in a sense that
they are ingredients of the superbridge.
%%%%%%%%%%%%%%%%%%%%%%%%%%%%%%%%%%%%%%%%%%%%%%%
%  というか，
%　「組換え理論では，(末端プリマリー鎖を通して)間接的にしか
%　取り入れることが出来ない」
%　ということ．
%%%%%%%%%%%%%%%%%%%%%%%%%%%%%%%%%%%%%%%%%%%%%%%%%
However, these chains can be treated in this theoretical framework 
only indirectly as in the previous theory.\cite{annable1}
%will be considered here only indirectly.
%%%%%%%%%%%%%%%%%%%%%%%%%%%%%%%%%%%%%%%%%%%%%%%%%
%Instead, 
%Therefore,
Effects of these internal chains are 
%effectively 
taken into account
via the end primary chains of the superbridge in the rest of this paper.
%
%For example, we assume that breakage rate of the end primary chains is larger than
%that of the primary bridge 
%
For example, 
the breakage rate of internal chains is reflected in the dissociation rate
of the end primary chains.
Furthermore, 
the elasticity of superbridges
%can be
is represented by that of the end primary chains
%under the assumption
by assuming that the end primary chains deform affinely
(see (\ref{dfepwf})).
This is valid when the macroscopic deformation applied to the gel 
is small as in the present case.
}
If $i_{\mbox{\scriptsize{A'}}}\ge3$, then the chain is a primary bridge,
so that the dissociation rate of the end group from the junction A' is $\beta$.
If $i_{\mbox{\scriptsize{A'}}}=2$, on the other hand, 
the chain is the end primary chain of the superbridge
(see Fig.\ref{super8}),
and we assume that 
%
%
%In this case, 
%in order to
%effects of 
the breakage rate of internal chains 
from the junctions B,B',C,...
%must be 
is reflected in the dissociation rate of this end primary chain
from the junction A'.
%since we are not 
%
%
%We assume 
%that the dissociation rate of the end primary chain 
%reflects effects of the internal chains must be reflected in 
%
Then we can put it as the sum of its own dissociation rate $\beta$
and the dissociation rate $2(s_b-1)\beta$ of $2(s_b-1)$ functional groups B,B',...
on the internal primary chains. 
As a result, the dissociation rate of A' {\it on average} can be expressed as
$\beta + 2(s_b-1)\rho\beta\equiv \beta^{\eff}$,
where $\rho$ is the probability for 
$i_{\mbox{\scriptsize{A'}}}$ to be $2$ and is given by
$\rho=m(i\ge3,\ip=2)/(m(i\ge3,\ip\ge3)+m(i\ge3,\ip=2))
=\nu^{\eff}_{pseud}/(\tilde{\nu}^{\eff}+\nu^{\eff}_{pseud})$.
We replace $\beta$ in (\ref{kretoek2}) with $\beta^{\eff}$ in the following
in order to incorporate the short lifetime of superbridges into account.
It should be noted that the equilibrium condition (\ref{eqcon}) still holds
after this replacement, so that the discussion given in \ref{subseceqi}
and \ref{subsecsubri} of this paper and IV in I remains valid.
When $\Delta$ is small, $\beta^{\eff}$ is inversely proportional to $\Delta$
since $s_b\sim1/\Delta$ while $\rho\sim1$ for $\Delta\ll1$.
Therefore, we see that the relaxation time $\tau$ of the gel
that is approximately given as the reciprocal of $\beta^{\eff}$
is proportional to $\Delta$ near the sol/gel transition concentration.

%%%%%%%%%%%%%%%%%%%%%%%%%%%%%%%%%%%%%%%%%%

Let $\tilde{P}_{k,\kp}^{total}$ be the probability for a ($k,\kp$)-chain
to be a primary bridge or an end primary chain of a superbridge.
It is given by
\bea
P_{k,\kp}^{total}=\frac{(\nu_{total}^{\eff})_{k,\kp}}{\nu_{k,\kp}},
\eea
where
\bea
(\nu_{total}^{\eff})_{k,\kp}
=\frac{\nu_{total}^{\eff}}{2}
\left(
\frac{\chi_k^{\eff}}{\nu_{SC}^{\eff}}
\frac{\tilde{\chi}_{\kp}^{\eff}}{\tilde{\nu}^{\eff}}
+\frac{\tilde{\chi}_k^{\eff}}{\tilde{\nu}^{\eff}}
\frac{\chi_{\kp}^{\eff}}{\nu_{SC}^{\eff}}
\right)
\label{dnumekrkp}
\eea
is the number of ($k,\kp$)-chains that is the primary bridge or
the end primary chain of the superbridge.
($\tilde{\chi}_k^{\eff}\equiv \sum_{i=2}^k\chi_{i,k}
=\chi_k(1-\zeta_0)(1-\zeta_0^{k-1})$ is the
number of paths ($\ge 2$) emersed from the $k(\ge 2)$-junction,
and hence $\tilde{\chi}_k^{\eff}/\tilde{\nu}^{\eff}$ is the probability
that a $k$-junction satisfies the condition $i\!\ge\!2$.
On the other hand, $\chi_k^{\eff}/\nu_{SC}^{\eff}$ 
is the probability for a $k$-junction to fulfill the condition $i\ge3$.
Eq. (\ref{dnumekrkp}) is expressed in a symmetric form.)
Eq. (\ref{dnumekrkp}) satisfies the following relation:
\bea
\sum_{k\ge 2}\sum_{\kp\ge 2}(\nu^{\eff}_{total})_{k,\kp}=\nu^{\eff}_{total}
\label{suntotal}
\eea
as it should be.\footnote{Note that $\chi_{2}^{\eff}/\nu_{SC}^{\eff}=0$.}
If a macroscopic deformation is applied to the gel,
not only primary bridges but also superbridges deform accordingly.
According to (\ref{nonafifine2}),
we assume that the rate of deformation vector of the ($k,\kp$)-chain is given by
\bea
\dot{\br}_{k,\kp}(t)=\tilde{P}_{k,\kp}^{total}\hat{\kappa}(t)\br.
\label{dfepwf}
\eea
By substituting (\ref{dfepwf}) into (\ref{baseeqige3}),
we can obtain the equations for $g^{\prime(\prime\prime)}_{k,\kp}$
that is given by (\ref{kretoek2}) but with $(\nu^{\eff}_{total})_{k,\kp}$
instead of $\nu^{\eff}_{k,\kp}$.
Therefore, in the high frequency limit,
$g^{\prime}_{k,\kp}$ reduces to $(\nu^{\eff}_{total})_{k,\kp}$.
The total (observable) modulus is given by
\bea
G^{\prime(\prime\prime)}(\omega)
=k_BT\sum_{k\ge 2}\sum_{\kp\ge 2}g_{k,\kp}^{\prime(\prime\prime)}(\omega),
\label{totmod}
\eea
and then we find $G^{\prime}\to 
G_{\infty}=k_BT \nu^{\eff}_{total}$ for $\omega\to\infty$.

%%%%%%%%%%%%%%%%%%%%%%%%%%%%%%%%%%%%%%%%%%%%%%%%%%%%%%%%%%%%%%%%%%%%%%%%%%%%%%%%%%
%  ・擬アフィン変形，
%　・端のprimary鎖に代表させる，
%　
%　ことについてレフェリーに言われたら，
%　
%　「これは物理的な仮定というよりも，技術上の問題であり，
%　　要するに，いかに組換え網目理論でどうすれば
%　　連結度やスーパーブリッジを取り扱うことができるのか，という技術の話である．」
%　
%　と言おう．
%
%  また，母集団として一端がi3以上，もう一端がi2以上の鎖を考えているが，
%　これで十分である．なぜなら，これらの鎖が弾性に寄与するのであり，
%　こいつの解離率の変化を知りたいのだから．
%　(弾性に寄与しないやつはどうでもよし)
%
%
%%%%%%%%%%%%%%%%%%%%%%%%%%%%%%%%%%%%%%%%%%%%%%%%%%%%%%%%%%%%%%%%%%%%%%%%%%%%%%%%%%

%%%%%%%%%%%%%%%%%%%%%%%%%%%%%%%%%%%%%%%%%%%%%%%%%%%%%%%%%%%%%%%%%%%%%%%%%%%%%%%%%%%
%%%%%%%%%%%%%%%%%%%%%%%%%%%%%%%%%%%%%%%%%%%%%%%%%%%%%%%%%%%%%%%%%%%%%%%%%%%%%%%%%%%
%%%%%%%%%%%%%%%%%%%%%%%%%%%%%%%%%%%%%%%%%%%%%%%%%%%%%%%%%%%%%%%%%%%%%%%%%%%%%%%%%%%
%%%%%%%%%%%%%%%%%%%%%%%%%%%%%%%%%%%%%%%%%%%%%%%%%%%%%%%%%%%%%%%%%%%%%%%%%%%%%%%%%%%
\section{Results and Discussions}
\label{resultsec}

\subsection{Saturating Junction Model}

\begin{figure}[t]
\begin{center}
\includegraphics*[scale=0.5]{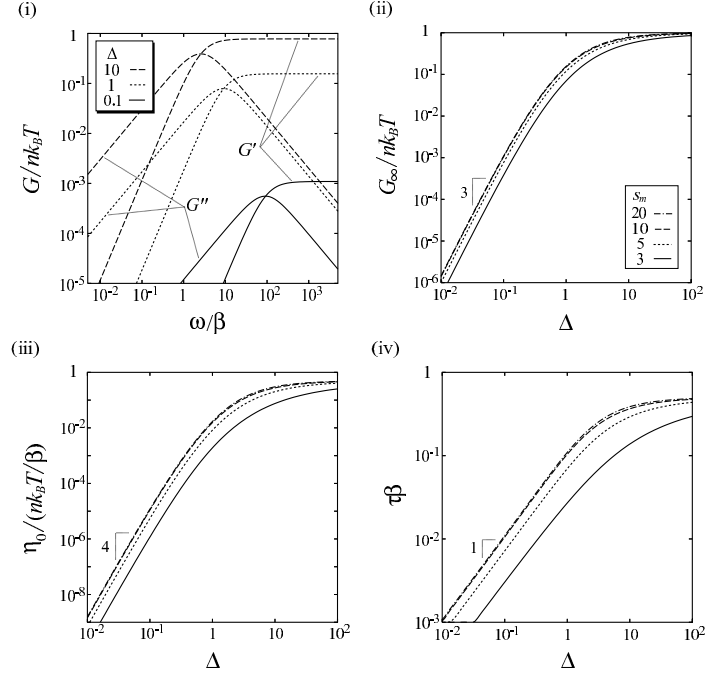}
\end{center}
\vspace*{-0.5cm}
\caption{
(i) The dynamic shear moduli (reduced by $nk_BT$) 
for the saturating junction model as a function of the frequency.
The relative concentration deviation $\Delta=(c-c^*)/c^*$ 
from the sol/gel transition concentration $c^*$ is varying from curve to curve,
while the maximum multiplicity of the junction is fixed at $s_m=20$.
(ii) The reduced plateau modulus, 
(iii) reduced zero-shear viscosity, 
and (iv) relaxation time
plotted against the relative concentration deviation
for several maximum multiplicities (increasing from bottom to top).
}
\label{satgetatau1}
\end{figure}

\begin{figure}[t]
\begin{center}
\includegraphics*[scale=0.5]{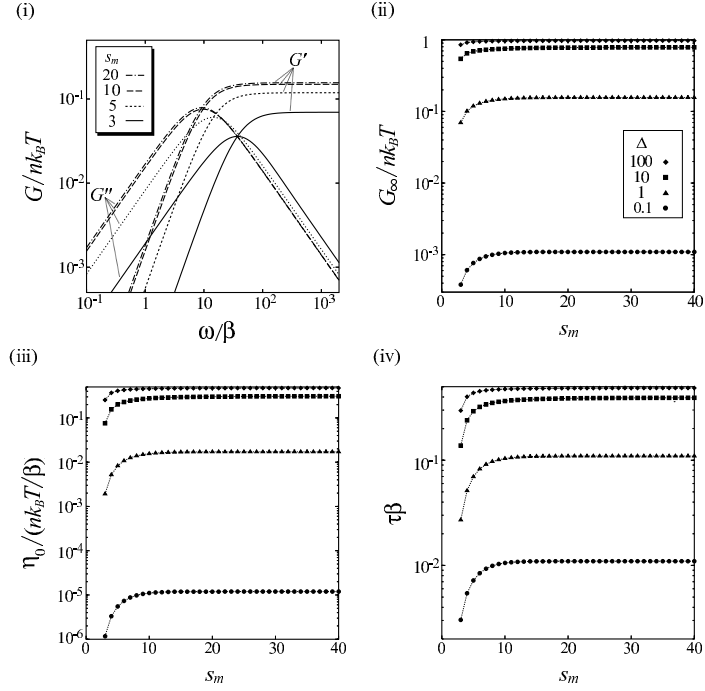}
\end{center}
\vspace*{-0.5cm}
\caption{
(i) The dynamic shear moduli (reduced by $nk_BT$) obtained 
for the saturating junction model as a function of the frequency.
The maximum multiplicity $s_m$ is varying from curve to curve 
with a relative concentration deviation fixed at $\Delta=1$.
(ii) The reduced plateau modulus, 
(iii) reduced zero-shear viscosity, and 
(iv) relaxation time
plotted against the maximum multiplicity
for several relative concentration deviation (increasing from bottom to top).
}
\label{satgetatau2}
\end{figure}

Fig.\ref{satgetatau1} (i) shows the dynamic shear moduli 
calculated from (\ref{totmod}) for the saturating junction model.
Relative concentration deviation is varying from curve to curve.
They are well described in terms of 
the Maxwell model with a single relaxation time.
Near the sol/gel transition concentration ($\Delta\ll1$),
the plateau modulus and the relaxation time increase as
$G_{\infty}\sim\Delta^3$ 
(see Fig.\ref{satgetatau1} (ii))
and $\tau\sim\Delta$
(Fig.\ref{satgetatau1} (iv)), respectively.
As a result, the zero shear viscosity increases as 
$\eta_0\sim G_{\infty}\tau\sim\Delta^4$
(Fig.\ref{satgetatau1} (iii)).
Note that these powers stem from
the mean-field treatment.\footnote{Rubinstein 
and Semenov have found the same power laws
from the mean-field treatment
for multifunctional polymers that can connect with each other
through pairwise association between functional groups on polymers.\cite{rubin2}}
For example, we can explain $\tau\sim\Delta$ as follows.
Let $\xi$ be the radius of gyration of the superbridge.
If we assume that the superbridge obeys the Gaussian statistics,
it is estimated to be $\xi\sim s_b^{1/2}$.
On the other hand, $\xi$ corresponds to the network mesh size,
and hence it obeys the scaling law $\xi\sim\Delta^{-\nu}$ 
with $\nu=1/2$ for $\Delta\ll 1$.
By comparing two expressions, we can find $s_b\sim 1/\Delta$.
Thus the mean lifetime of bridges
(primary bridges and superbridges) 
that corresponds to the relaxation time of the network
is approximately estimated to be
$\tau\sim 1/[\beta+2(s_b-1)\rho\beta]\sim \Delta/\beta$.
Fig.\ref{satgetatau2} shows the dynamic shear moduli 
as a function of the maximum multiplicity $s_m$.
The relaxation time increases with $s_m$
because the number of superbridges decreases as $s_m$ increases.

The reduced plateau modulus explicitly
depends only on the reduced polymer concentration and
the maximum multiplicity of the junction.
%($s_m$ in the case of the saturating junction model).
Therefore, it is written as
\bea
\frac{G_{\infty}}{nk_BT}=f_1(c,s_m).
\label{gscale2}
\eea
Similarly, the reduced zero-shear viscosity and the relaxation time
can be expressed as
\bea
\frac{\beta\eta_0}{nk_BT}=f_2(c,s_m)
\label{etascale2}
\eea
and 
\bea
\beta\tau=f_3(c,s_m),
\label{tauscale2}
\eea
respectively
($f_1\sim f_3$ are dimensionless functions of $c$ and $s_m$).
In order to investigate how the dynamic shear moduli depend on the temperature 
and the polymer volume fraction, let us rewrite
(\ref{gscale2}) $\sim$ (\ref{tauscale2}) as
\bea
& &
\frac{v_0G_{\infty}}{k_BT_0}=\frac{\phi}{N}\frac{T}{T_0}f_1(c(T/T_0,N,\phi),s_m), \\
\label{reresceg}
& &
\frac{v_0\beta_0\eta_0}{k_BT_0}=\frac{\phi}{N}\frac{T}{T_0}e^{T_0/T-1}
f_2(c(T/T_0,N,\phi),s_m), \label{reresceeta}\\
& &\beta_0\tau=e^{T_0/T-1}f_3(c(T/T_0,N,\phi),s_m),
\label{rerescetau}
\eea
respectively.
We have put $W=\epsilon$ in the derivation of (\ref{reresceeta}) and (\ref{rerescetau}),
In this case, the dissociation rate at temperature $T$ is written as
\bea
\beta=\omega_0 e^{-\epsilon/k_BT}=\beta_0 e^{1-T_0/T}
\label{betahito}
\eea
where $\beta_0$ is the dissociation rate at temperature $T_0$.
Recall that the reduced concentration depends on the temperature,
molecular weight, and the polymer volume fraction as
$c(T/T_0,N,\phi)=2\phi\lambda_0e^{T_0/T}/N$.
Thus, for example, the zero-shear viscosity 
(\ref{reresceeta}) depends on the temperature
through $\beta$, $c$ and a prefactor.
\begin{figure}[t]
\begin{center}
\includegraphics*[scale=0.5]{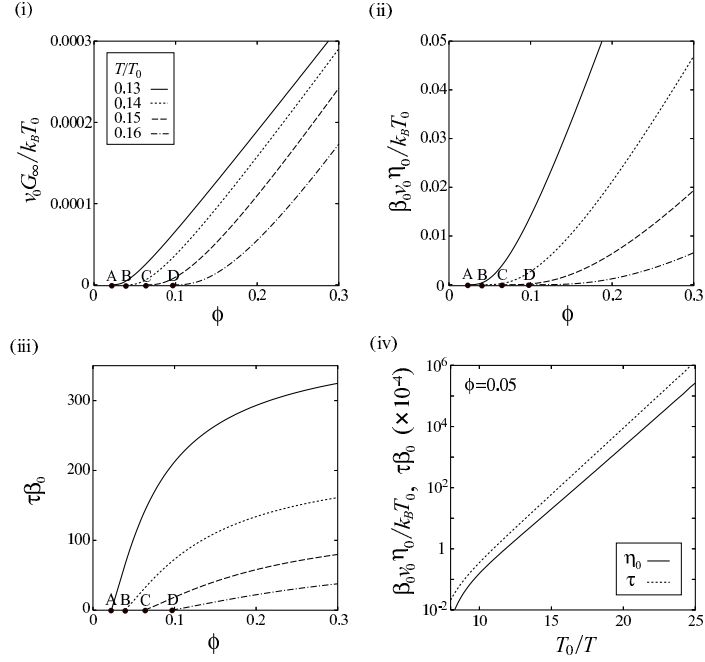}
\end{center}
\vspace*{-0.5cm}
\caption{
(i) The plateau modulus,
(ii) zero-shear viscosity, and
(iii) relaxation time plotted against the polymer volume fraction
for several temperature with $s_m\!=\!20$, $N\!=\!100$ and $\lambda_0\!=\!1$.
Marked points A, B, C, D 
on the horizontal axis
of (i) indicate the critical volume fraction $\phi^{*}$
for each temperature
corresponding to the points in Fig.\ref{solge} (b).
(iv) Arrhenius plots of the 
zero-shear viscosity and the 
relaxation time with $\phi\!=\!0.05$, $s_m\!=\!20$, $N\!=\!100$ and $\lambda_0\!=\!1$.
}
\label{satgeta-T}
\end{figure}
The unitless plateau modulus (\ref{reresceg}), 
zero-shear viscosity (\ref{reresceeta})
and relaxation time (\ref{rerescetau})
are shown in Fig.\ref{satgeta-T} (i) $\sim$ (iii)
as a function of the polymer volume fraction for several temperature.
Volume fraction is varying
across the sol/gel transition line drawn in Fig.\ref{solge} (b)
for each temperature.
If the volume fraction $\phi$ is small,
$G_{\infty}$ and $\eta_0$ depend on $\phi$ through the reduced concentration $c$
and a prefactor, while they are approximately
proportional to $\phi$ if $\phi$ (and hence $c$) is large because
$f_1$ and $f_2$ depend only weakly on $c$ (see Fig.\ref{satgetatau1})
in this case.
On the other hand, $\tau$ depends on $\phi$ only through $c$.
As shown in Fig.\ref{satgeta-T} (iv), 
%shows the Arrhenius plot of 
the zero-shear viscosity and the relaxation time
approximately show the Arrhenius law temperature dependences.
At higher temperature, 
we can see a slight deviation from the Arrhenius law.
This deviation stems from the fact that $\eta_0$ and $\tau$ depend on $T$
not only through $\beta$ (see (\ref{betahito}))
but also through $c$ (and a prefactor in the case of $\eta_0$),
On the other hand, $\eta_0$ and $\tau$
depend only weakly on $c$ at lower temperature,
and hence they show approximately the Arrhenius law.
We can guess from (\ref{reresceg}) and (\ref{rerescetau}) that
the dynamic shear moduli at temperature $T$ can be superimposed to
the curve at the reference temperature $T_{ref}$
if they are horizontally and vertically shifted by a factor of $a_T$ 
and $b_T$, respectively, where
\begin{subequations}\label{bkshift}
\bea
& &a_T=\exp\left[-T_0\left(\frac{1}{T_{ref}}-\frac{1}{T}\right)\right]
\cdot\frac{f_3(c(T/T_0,N,\phi),s_m)}{f_3(c(T_{ref}/T_0,N,\phi),s_m)}, \\
& &b_T=\frac{T_{ref}}{T}
\cdot\frac{f_1(c(T_{ref}/T_0,N,\phi),s_m)}{f_1(c(T/T_0,N,\phi),s_m)}.
\eea
\end{subequations}
Especially for larger $\phi$ or lower $T$, 
(\ref{bkshift}) is approximately written as
\begin{subequations}\label{anashift}
\bea
& &a_T\simeq\exp\left[-T_0\left(\frac{1}{T_{ref}}-\frac{1}{T}\right)\right], \\
& &b_T\simeq\frac{T_{ref}}{T}
\eea
\end{subequations}
It has been revealed by Annable {\it et al.} 
that the shift factor given by
(\ref{anashift}) produces the master curve successfully.\cite{annable1}

\begin{figure}[t]
\begin{center}
\includegraphics*[scale=0.5]{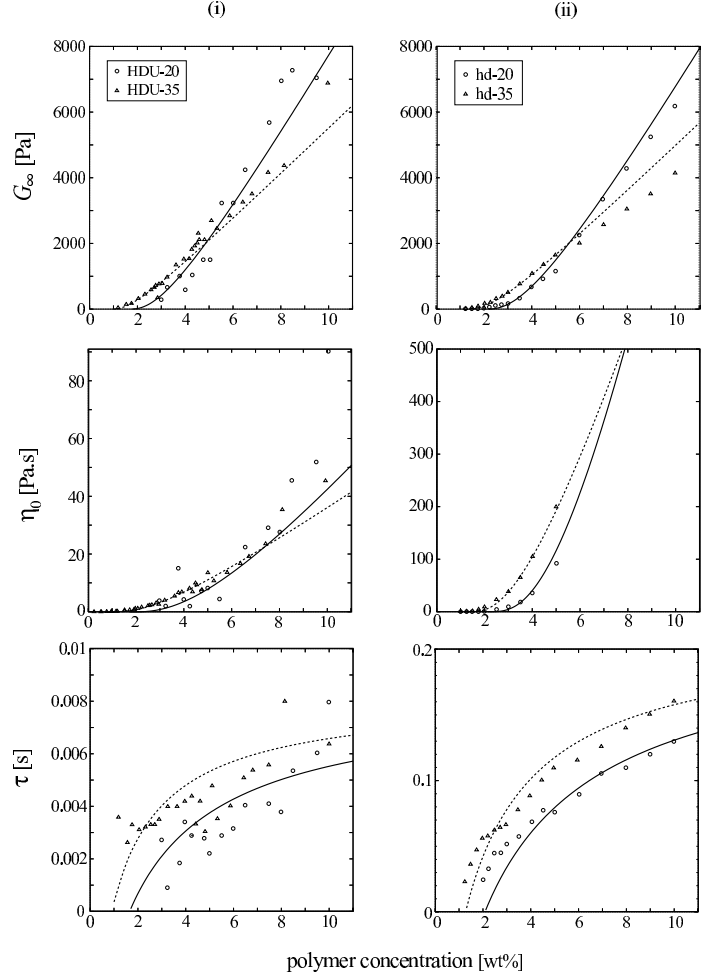}
\end{center}
\vspace*{-0.5cm}
\caption{
Comparison between the theoretically obtained
plateau modulus (top), zero-shear viscosity (middle), relaxation time (bottom)
for the saturating junction model (lines)
and experimental data obtained 
for (i) telechelic PEO with narrow molecular weight distribution and
fully end-capped with C$_{16}$ alkanes
($M_w$=20,000 for HDU-20 \cite{meng} and $M_w$=35,000 for HDU-35 \cite{pham})
and (ii) HEUR 
end-capped with the same alkanes
($M_w$=20,000 for hd-20 and $M_w$=33,100 for hd-35 \cite{annable1}).
Values of molecular parameters $s_m$, $\beta$ and $\lambda v_0$ 
used to draw theoretical curves are listed in TABLE \ref{table1}.
}
\label{exphikaku1}
\end{figure}

In Fig.\ref{exphikaku1} (i),
we compare theoretically predicted 
dynamic shear moduli (i.e., plateau modulus, zero-shear viscosity 
and relaxation time)
with experimental observation for aqueous solutions of 
telechelic PEO of 20 kg/mol \cite{meng} and 35 kg/mol \cite{pham}
with narrow-molecular-weight distribution
and fully end-capped with C$_{16}$ alkanes.
We call these polymers HDU-20(35) according to ref..\cite{meng}
The reduced concentration $c$ was converted
into the polymer concentration in weight percentage $c_w$ 
through a relation $c=(2000N_A/M)\lambda v_0 c_w$ 
($N_A$ is Avogadro's number).
We have three molecular parameters for a given molecular weight: 
$s_m$, $\lambda v_0$ and $\beta$. 
(Note that $\beta$ is not required to calculate $G_{\infty}$.)
Values of these parameters used to draw theoretical curves 
are listed in TABLE \ref{table1}.
We 
%now 
find better agreements between theory and experiment 
than in the case that 
the short lifetime of 
superbridges 
is not taken into consideration.\cite{in1}
The value of $\lambda v_0$ increases with increasing the molecular weight.
%%%%%%%%%%%%%%%%%%%%%%%%%%%%%%%%%%%%%%%%%%%%%%%%%%%%%%%%%%%%%%%%%%%%
%
%　ここの物理的な理由を聞かれるかもしれない．
%　v_0が大きいということは，くっつきやすいということ．
%　会合数が同じで鎖が長いほど，，コアにくっつきやすくなるとはこれいかに．
%　わからないときは，「わからない」と書くべき．
%
%%%%%%%%%%%%%%%%%%%%%%%%%%%%%%%%%%%%%%%%%%%%%%%%%%%%%%%%%%%%%%%%%%%%
This indicates that the effective volume $v_0$ of a functional group
increases with increasing the chain length.
In Fig.\ref{exphikaku1} (ii), we attempt to fit theoretical curves 
to experimental data reported by Annable {\it et al.} \cite{annable1} for
HEUR of the similar molecular weight (but with broader molecular weight distribution)
and end-capped with C$_{16}$ alkanes. 
They are called hd-20(35) after ref..\cite{meng}
Parameter values adopted to fit experimental data 
are also listed in TABLE \ref{table1}.
We still find a good agreement between theory and experiment
in spite of a broader molecular weight distribution of hd polymers.
A difference in the value of $\lambda v_0$ between
HDU and hd for each (averaged) molecular weight
might stem from a difference in the polydispersity of the PEO backbone.
A ratio between the value of $\lambda v_0$ for HDU-20 and for HDU-35
(3.2) is close to 
%the ratio 
that
between hd-20 and hd-35 (2.9).
A discrepancy in the value of $\beta$ for HDU and for hd
might stem from the difference
in the coupling agents between the alkanes and the PEO backbone 
as suggested in refs..\cite{pham,meng}

\begin{table}[t]
\begin{center}
\begin{tabular}{lccc} 
\hline\hline
polymer ~~ & ~~$s_m$~~  & ~~$\beta$ [1/s]~~ & ~~$\protect\lambda v_0\times 10^{23}$ [m$^3$]~~ \\ \hline
HDU-20 &  20 & 60  & 0.1 \\ %\hline
HDU-35 &  20 & 60  & 0.32 \\ %\hline
hd-20  &  20 & 2.3 & 0.08 \\ %\hline
hd-35  &  20 & 2.3 & 0.23 \\ \hline\hline
\end{tabular}
\end{center}
\caption{Values of molecular parameters
used in Fig.\ref{exphikaku1}.
}
\label{table1}
\end{table}

%%%%%%%%%%%%%%%%%%%%%%%%%%%%%%%%%%%%%%%%%%%%%%%%%%

\subsection{Fixed Multiplicity Model}

\begin{figure}[t]
\begin{center}
\includegraphics*[scale=0.5]{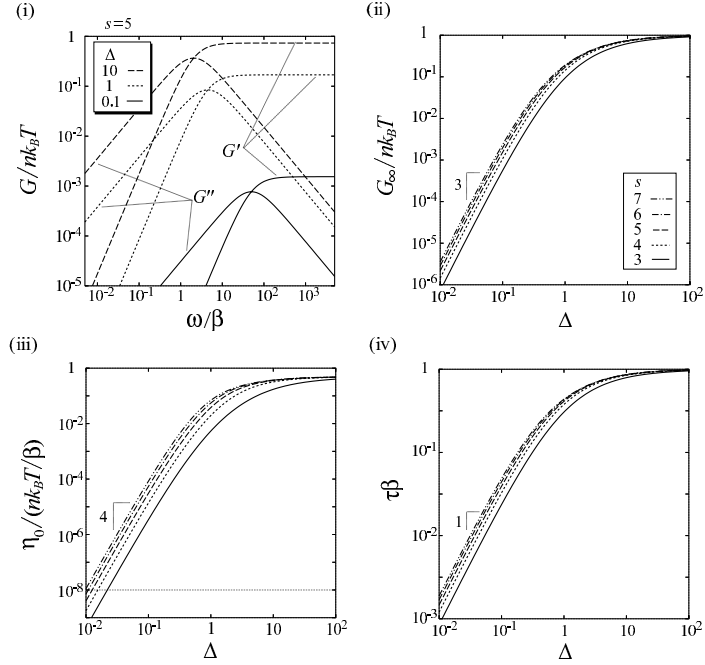}
\end{center}
\vspace*{-0.5cm}
\caption{
(i) The dynamic shear moduli (reduced by $nk_BT$) obtained 
for the fixed multiplicity model as a function of the frequency.
The relative concentration deviation 
is varying from curve to curve,
while the multiplicity of the junction is fixed at $s=5$.
(ii) The reduced plateau modulus, 
(iii) reduced zero-shear viscosity, and 
(iv) relaxation time
plotted against the relative concentration deviation 
for several multiplicities.
}
\label{fixgetatau1}
\end{figure}

Fig.\ref{fixgetatau1} shows the dynamic shear moduli
of the fixed multiplicity model 
%($\delta=0.01$)
together with the plateau modulus, zero-shear viscosity and relaxation time
plotted against the relative concentration deviation.
These quantities obey the same critical behavior as in the saturating junction model,
i.e., they increase as 
$G_{\infty}\sim\Delta^3$, $\eta_0\sim\Delta^4$ and $\tau\sim\Delta$
with increasing $\Delta$ near the sol/gel transition concentration.

\begin{figure}[t]
\begin{center}
\includegraphics*[scale=0.5]{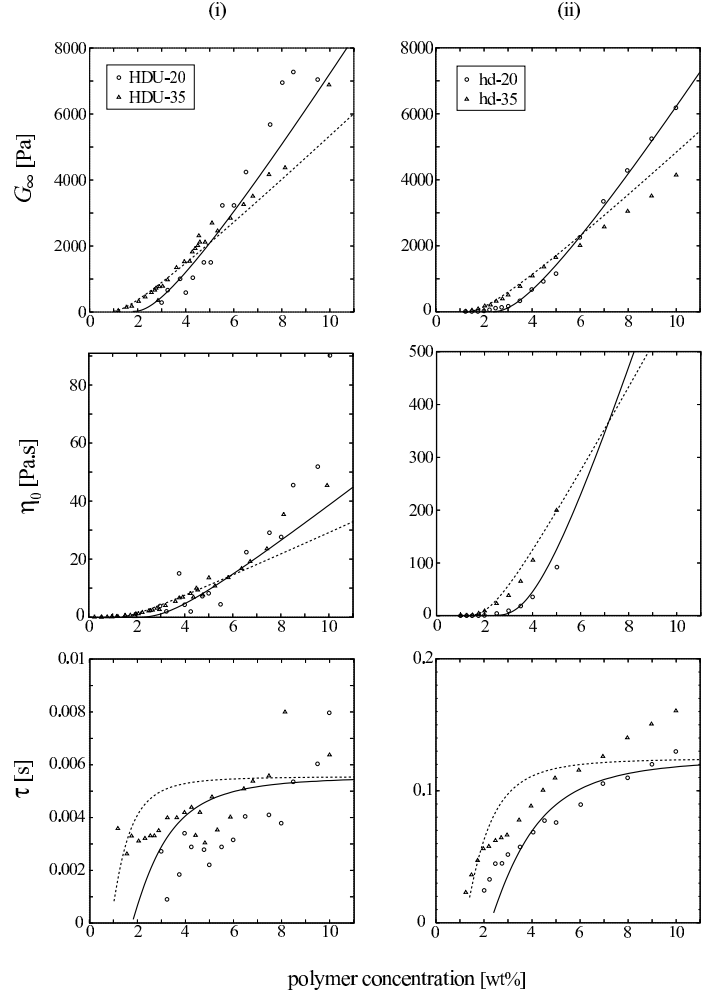}
\end{center}
\vspace*{-0.5cm}
\caption{
Comparison between the theoretically predicted 
plateau modulus (top), zero-shear viscosity (middle), relaxation time (bottom)
for the fixed multiplicity model (lines)
and experimental data observed 
for (i) telechelic PEO with narrow molecular weight distribution
fully end-capped with C$_{16}$ alkanes
($M_w$=20,000 for HDU-20 \cite{meng} and $M_w$=35,000 for HDU-35 \cite{pham})
and (ii) HEUR 
end-capped with the same alkanes
($M_w$=20,000 for hd-20 and $M_w$=33,100 for hd-35 \cite{annable1}).
Values of molecular parameters $s$, $\beta$ and $\lambda v_0$ 
used to draw theoretical curves are listed in TABLE \ref{table2}.
}
\label{exphikaku2}
\end{figure}

\begin{table}[t]
\begin{center}
\begin{tabular}{lccc} 
\hline\hline
polymer ~~ & ~~$s$~~  & ~~$\beta$ [1/s]~~ & ~~$\lambda v_0\times 10^{23}$ [m$^3$]~~ \\ 
\hline
HDU-20 &  6 & 90 & 0.09 \\ %\hline
HDU-35 &  6 & 90 & 0.32 \\ %\hline
hd-20  &  6 & 4  & 0.07 \\ %\hline
hd-35  &  6 & 4  & 0.23 \\ \hline\hline
\end{tabular}
\end{center}
\caption{
Parameter values used in Fig.\ref{exphikaku2}.
}
\label{table2}
\end{table}

In Fig.\ref{exphikaku2}, theoretical curves are compared with 
experimental data for telechelic PEO.
Values of parameters used to draw theoretical curves 
are listed in TABLE \ref{table2}.
We find a disagreement between
theory and experiment in the relaxation time;
theoretical curves increase more rapidly 
with increasing the concentration than experimental data,
and they become approximately flat above a certain concentration.
This is because the fraction of junctions with the path connectivity $i=2$
(or, in other words, the number of superbridges) is small
when the junction can take only a single multiplicity.\footnote{The relaxation time 
does not depend on the concentration for the fixed multiplicity model 
if effects of superbridges are not taken into account.\cite{in1}}
This tendency becomes more pronounced with increasing the multiplicity
because the fraction of $i=2$ junctions is smaller for larger multiplicity.
Thus we guess that the multiplicity should not be fixed at a single value.
In real systems, junctions might be cores of flower-like micelles
and the aggregation number (i.e., the number of chains per junction), 
say $s_{flower}$, 
is almost independent of the polymer concentration
as some researchers indicated.
In this case, the number of bridge and dangling chains emersed from
the junction (i.e., multiplicity of the present theory) 
can be less than $s_{flower}$.
This situation corresponds to the saturating junction model, 
not the fixed multiplicity model. 
This might be the reason why the saturating junction model
can describe the dynamic shear moduli of 
%the real system 
telechelic PEO
better than the fixed multiplicity model.
%as shown in Fig.\ref{exphikaku1}.

%%%%%%%%%%%%%%%%%%%%%%%%%%%%%%%%%%%%%%%%%%%%%%%%%%%%%%%%%%%%%%%%%%%%%%%%%%%%%
%%%%%%%%%%%%%%%%%%%%%%%%%%%%%%%%%%%%%%%%%%%%%%%%%%%%%%%%%%%%%%%%%%%%%%%%%%%%%
%%%%%%%%%%%%%%%%%%%%%%%%%%%%%%%%%%%%%%%%%%%%%%%%%%%%%%%%%%%%%%%%%%%%%%%%%%%%

\section{Summary}
\label{conlsec}

We developed the theory of transient networks
with junctions of limited multiplicity.
The global information was incorporated into the theory
by introducing the elastically effective chains (active chains) 
according to the criterion by Scanlan and Case
and by considering the effect of 
superbridges whose backbone is formed by several chains connected in series.
Linear viscoelasticities of the network 
were studied as functions of thermodynamic quantities.
Near the critical concentration for the sol/gel transition,
superbridges are infinitely long along the backbone and 
their number is comparable with that of primary bridges.
Thus the mean lifetime of bridges is quite short near the critical point
and so does the relaxation time.
It was found that the relaxation time is proportional to the 
concentration deviation $\Delta$ near the sol/gel transition concentration.
Since the plateau modulus increases as the cube of $\Delta$
as a result of the mean-field treatment,
the zero-shear viscosity increases as $\Delta^4$ near the gelation point.
Obtained dynamic shear moduli 
as a function of the polymer concentration were found to agree well with
the experimental data observed for aqueous solutions of telechelic 
poly(ethylene oxide).

We assumed in this theoretical model that 
intramolecular associations generating looped chains are absent.
Looped chains are supposed to compete with 
intermolecular association that causes bridge chains
at a junction due to the limitation of the multiplicity that the junction can take.
Such competition might influence the viscoelasticity of the system.
This effect as well as the influence 
of additives such as surfactant or single end-capped polymers
will be studied in the forthcoming paper.

%%%%%%%%%%%%%%%%%%%%%%%%%%%%%%%%%%%%%%%%%%%%%%%%%%%%%%%%%%%%%%%%%%%%%%%%%%%%%
%%%%%%%%%%%%%%%%%%%%%%%%%%%%%%%%%%%%%%%%%%%%%%%%%%%%%%%%%%%%%%%%%%%%%%%%%%%%%
%%%%%%%%%%%%%%%%%%%%%%%%%%%%%%%%%%%%%%%%%%%%%%%%%%%%%%%%%%%%%%%%%%%%%%%%%%%%

\bibliographystyle{aipprocl}

\begin{thebibliography}{999}



\bibitem{winyek}
M. A. Winnik, and A. Yekta,
{\bf 2}, 424 (1997).


\bibitem{annable1}
T. Annable, R. Buscall, R. Ettelaie, and D. Whittlestone,
J. Rheol.,
{\bf 37}, 695 (1993).


\bibitem{jen2}
R. D. Jenkins, D. R. Bassett, C. A. Silebi, and M. S. El-Aasser,
J. Appl. Polym. Sci.,
{\bf 58}, 209 (1995).


\bibitem{winnik}
A. Yekta, B. Xu, J. Duhamel, H. Adiwidjaja, M. A. Winnik,
Macromolecules,
{\bf 28}, 956 (1995).


\bibitem{fran1}
E. Alami, M. Almgren, W. Brown, and J. Franc\c{o}is,
Macromolecules,
{\bf 29}, 2229 (1996).


\bibitem{pham0}
Q. T. Pham, W. B. Russel, J. C. Thibeault, and W. Lau,
Macromolecules,
{\bf 32}, 2996 (1999).


\bibitem{pham}
Q. T. Pham, W. B. Russel, J. C. Thibeault, and W. Lau,
Macromolecules,
{\bf 32}, 5139 (1999).


\bibitem{serero0}
Y. S\'{e}r\'{e}ro, V. Jacobsen, J. -F. Berret, and R. May,
Macromolecules 
{\bf 33}, 1841 (2000).


\bibitem{serero}
D. Calvet, A. Collet, M. Viguier, J. -F. Berret, and Y. S\'{e}r\'{e}ro,
Macromolecules,
{\bf 36}, 449 (2003).


\bibitem{kujawa}
P. Kujawa, H. Watanabe, F. Tanaka, F. M. Winnik,
Eur. Phys. J. E,
{\bf 17}, 129 (2005).


\bibitem{meng}
X. -X. Meng, W. B. Russel,
J. Rheol.,
{\bf 50}, 189 (2006).


\bibitem{in1}
T. Indei,
submitted for publication (2006).


\bibitem{scanlan}
J. Scanlan,
J. Polym. Sci.,
{\bf 43}, 501 (1960).


\bibitem{case}
L. C. Case,
J. Polym. Sci.,
{\bf 45}, 397 (1960).


\bibitem{tanastock}
F. Tanaka, and W. H. Stockmayer,
Macromolecules,
{\bf 27}, 3943 (1994).


\bibitem{tanaishi}
F. Tanaka, and M. Ishida,
Macromolecules,
{\bf 29}, 7571 (1996).


\bibitem{graessley}
D. S. Pearson, and W. W. Graessley,
Macromolecules,
{\bf 11}, 528 (1978).


\bibitem{flory}
P. J. Flory,
{\it Principles of Polymer Chemistry},
Cornell University Press: Ithaca, NY, 1953; Chapter 9.
%(本でなく，論文載せるか？)


\bibitem{stocky12}
W. H. Stockmayer,
J. Chem. Phys.,
{\bf 11}, 45 (1943);
Ibid. 
{\bf 12}, 125 (1944). 


\bibitem{tanaed1}
F. Tanaka, and S. F. Edwards,
Macromolecules,
{\bf 25}, 1516 (1992).


\bibitem{tanaed2}
F. Tanaka, and S. F. Edwards,
J. Non-Newtonian Fluid Mech.,
{\bf 43}, 247, 273, 289 (1992).


\bibitem{intana1}
T. Indei, F. Tanaka,
J. Rheol., 
{\bf 48}, 641 (2004).


\bibitem{fukuyama}
K. Fukui, and T. Yamabe,
Bull. Chem. Soc. Jpn.,
{\bf 40}, 2052 (1967).


\bibitem{rubin2}
M. Rubinstein, and A. N. Semenov,
Macromolecules,
{\bf 31}, 1386 (1998).


\end{thebibliography}

\end{document}